\documentclass{PoS}

\usepackage{amsmath,amsthm,amssymb}

\def\be{\begin{equation}}
\def\ee{\end{equation}}
\def\ba{\begin{align}}
\def\ea{\end{align}}

\def\lsim{\raise0.3ex\hbox{$\;<$\kern-0.75em\raise-1.1ex\hbox{$\sim\;$}}}
\def\gsim{\raise0.3ex\hbox{$\;>$\kern-0.75em\raise-1.1ex\hbox{$\sim\;$}}}
\def\eps{\varepsilon}
\def\theta{\vartheta}

\title{Numerical Simulations of Cosmic-Ray Acceleration at Core-Collapse Supernovae}

\ShortTitle{CR Acceleration at Core-Collapse Supernovae}

\author{\speaker{Gwenael Giacinti}\\
        Max-Planck-Institut f\"ur Kernphysik, Postfach 103980, 69029 Heidelberg, Germany\\
        E-mail: \email{giacinti@mpi-hd.mpg.de}}

\author{Vikram Dwarkadas\\
        Department of Astronomy and Astrophysics, University of Chicago, 5640 S Ellis Ave, Chicago, IL 60637, USA\\
        E-mail: \email{vikram@astro.uchicago.edu}}

\author{Alexandre Marcowith\\
        Laboratoire Univers et Particules de Montpellier (LUPM) Universit\'e Montpellier, CNRS/IN2P3, CC72, place Eug\`ene Bataillon, 34095, Montpellier Cedex 5, France\\
        E-mail: \email{Alexandre.Marcowith@univ-montp2.fr}}

\author{Andrea Chiavassa\\
        Universit\'e C\^{o}te d'Azur, Observatoire de la C\^{o}te d'Azur, CNRS, Lagrange, CS 34229, Nice, France\\
        E-mail: \email{andrea.chiavassa@oca.eu}}

\abstract{Core-collapse supernovae exploding in dense winds are favorable sites for cosmic-ray (CR) acceleration to very high energies. We present our CR-radiation-hydrodynamics simulations of the explosion of a red supergiant. We study the evolution of the shock wave during the first day following core collapse, and estimate the time at which CR acceleration can start. We then calculate the maximum CR energy at the forward shock as a function of time, and show that it may already exceed 100\,TeV only a few hours after shock breakout from the surface of the star.}

\FullConference{36th International Cosmic Ray Conference -ICRC2019-\\
		July 24th - August 1st, 2019\\
		Madison, WI, U.S.A.}

\begin{document}

\section{Introduction}
\label{Sec_Introduction}

Several studies have suggested that core-collapse supernovae (SN) exploding in dense winds could accelerate cosmic-rays (CR) to PeV energies for a few decades~\cite{Tatischeff:2009kh,Bell:2013kq,Marcowith:2014,Marcowith:2018ifh}. The situation during the first few days of the SN is somewhat more complicated. After core collapse, a radiation-mediated shock travels through the progenitor. Such a shock is not expected to accelerate CRs because its width is larger than the gyroradius of suprathermal particles. Later, at shock breakout (SB), it stalls in the outer layers of the star at an optical depth $\tau \sim c/U_{\rm s}$, where $U_{\rm s}$ denotes the shock velocity. At this point, the radiation in the immediate shock downstream escapes, producing a flash of photons~\cite{Colgate74}, which accelerates the circumstellar wind. A collisionless shock (CS) later forms~\cite{ChevalierKlein79}: The dilution of photons as $1/R^{2}$, where $R$ is the stellar radius ($R=0$ in the centre of the star), ensures that the shocked outer layers of the star ram supersonically into the wind further out. Once the conditions are favourable for the first order Fermi mechanism to operate at the newly formed CS, CR acceleration should start.

In this work, we focus on the first day of the explosion of a red supergiant, and study the beginning of CR acceleration in such an extreme environment. In Sect.~\ref{Sec_Simulations}, we present our simulations of the progenitor and of its explosion, and describe our calculations of the quantities related to particle acceleration. We present our results in Sect.~\ref{Sec_Results} and conclude in Sect.~\ref{Sec_Conclusions}.

\section{Numerical simulations}
\label{Sec_Simulations}

Stellar convection modelling is carried out using three-dimensional (3D) radiative hydrodynamical (RHD) code CO$^5$BOLD \cite{2012JCoPh.231..919F}. It solves the coupled non-linear equations of compressible hydrodynamics and non-local radiative energy transfer in the presence of a fixed external gravitational field. The computational domain is a cubic grid equidistant in all directions, and the same open boundary condition is employed for all sides of the computational box (see, e.g., Ref.~\cite{2011A&A...535A..22C}). The code employs realistic input physics either for the equation of state~\cite{2012JCoPh.231..919F} or for the radiative transfer opacities~\cite{2008A&A...486..951G}. A detailed and precise solution of the radiative transfer is essential for a realistic treatment of convection because it is the radiative losses in the surface layers of the star that drain the convective movements and thus influence the whole simulation domain.

\begin{figure*}
\begin{center}
\includegraphics[width=0.44\textwidth]{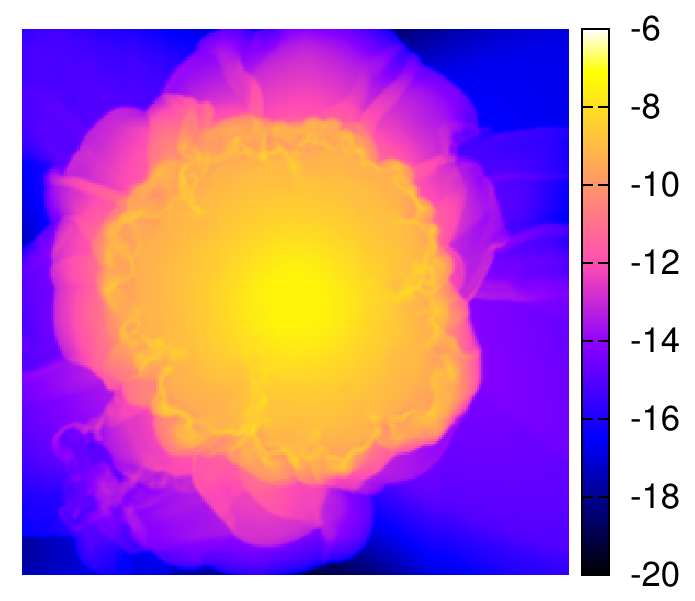}
\includegraphics[width=0.54\textwidth]{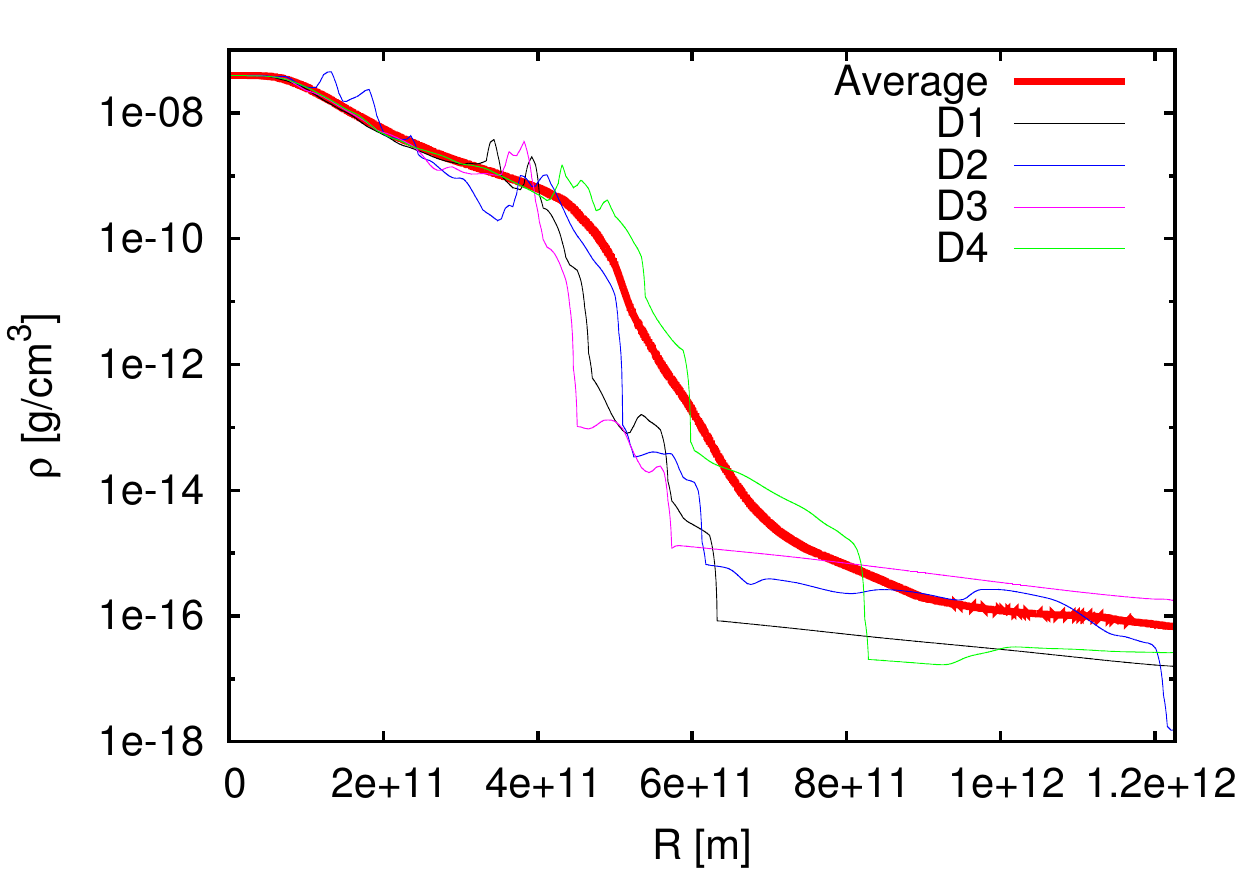}
\caption{\label{Fig1} (Initial) density profile of the progenitor red supergiant star considered in this study, and of its surroundings. Left panel: $\log \left( \rho/({\rm g \cdot cm}^{-3}) \right)$ in a plane containing the centre of the star $R=0$, located in the centre of the panel. See the colour code in the bar on the right-hand side. Right panel: Corresponding radial density profile averaged over all radial directions (thick red line), and density profiles along four given radial directions, \lq\lq D1\rq\rq\/ to \lq\lq D4\rq\rq\/ (thin lines, see the key).}
\end{center}
\end{figure*}

The fundamental stellar parameters of the RHD simulation used in this work are reported in Table~1 of Ref.~\cite{2018A&A...610A..29K}. We plot in Fig.~\ref{Fig1} a snapshot of the resulting density profile. In the left panel, we plot the logarithm of the density in a plane cut of the 3D simulated domain. The plane contains the centre of the star, $R=0$, in the centre of the panel. The very dense central region of the star is not represented. The dense, optically thick layers of the progenitor are in yellow/orange, while the surrounding dark regions correspond to the optically thin circumstellar material. Deviations from spherical symmetry are clearly visible. In the right panel, we plot with the thick red line the radial density profile of this star, averaged over all radial directions. The thin lines, denoted \lq\lq D1\rq\rq\/ to \lq\lq D4\rq\rq\/, represent the density profiles along four given radial directions.

The simulation of the SN is carried out using our Eulerian 1D-spherical radiation-hydro-dynamics code, presented in Ref.~\cite{GGABPaper1}. We use the average radial density and temperature profiles of the above star as an input. CRs have been added as a pressure term in our code, see Ref.~\cite{GGetal_In_Prep} for more details. The code is two-temperature, i.e. electron and proton temperatures are assumed to be equal. For the radiation, we use a gray frequency average, and represent it by its internal energy $E_{\rm rad}$ with characteristic temperature $T_{\rm rad} = ( cE_{\rm rad} / 4\sigma )^{1/4}$, where $\sigma$ denotes the Stefan-Boltzmann constant. The radiation transport is solved using a \lq\lq square-root\rq\rq\/ flux-limited diffusion approximation. We take into account Compton cooling and bremsstrahlung for the transfer of energy between fluid and radiation, using the formulae from Ref.~\cite{ChevalierKlein79}. At $t=0$ in the simulation, we trigger the SN by depositing $10^{51}$\,erg at the centre of the star as a spike in temperature. We follow the evolution of the physical conditions around the shock in the simulation, and determine the formation time of the CS, using the method of Ref.~\cite{GGABPaper1}. Once the CS is formed, we calculate, at each time in the simulation, the typical Coulomb loss time of suprathermal particles at the CS, $\tau_{\rm Coul, T_{\rm e,u}}$, and their typical acceleration time to 1\,GeV, $\tau_{\rm acc, 1GeV}$. For the latter, we assume Bohm diffusion of the CRs, in an {\it initial} circumstellar magnetic field of 1\,mG strength. For stronger magnetic fields, particle acceleration would proceed faster. We also calculate the characteristic times for CR energy losses due to inelastic $pp$ collisions, $\tau_{\rm pp}$, and adiabatic losses, $\tau_{\rm adiab}$. For the latter, we take the conservative lower value $\tau_{\rm adiab} = R_{\rm s}/U_{\rm s}$, where $R_{\rm s}$ denotes the radius of the CS. See Refs.~\cite{GGABPaper1,GGetal_In_Prep} for more details. We assume that particle acceleration starts in our simulation once the acceleration time to 1\,GeV is shorter than all loss times.

We assume an $E^{-2}$ spectrum for the CRs accelerated at the CS. We further assume that magnetic field amplification at the shock is due to Bell's instability~\cite{Bell2004}, which is driven by the escape of the highest energy CRs in the upstream of the CS~\cite{Bell:2013kq}. We calculate the maximum CR energy at the CS, $E_{\max}$, using the results of Ref.~\cite{Bell:2013kq}. At each time step in the simulation, we calculate the CR current escaping ahead of the CS, the growth rate of the instability in the upstream, and $E_{\max}$. See Ref.~\cite{GGetal_In_Prep} for technical details.

Finally, we calculate the CR acceleration time to $E_{\max}$ in the {\it amplified} magnetic field from Bell's instability, $\tau_{\rm acc, E_{\max}}$, and verify that it does not exceed the characteristic CR loss times $\tau_{\rm adiab}$, $\tau_{\rm pp}$, and $\tau_{\rm p\gamma}$, where $\tau_{\rm p\gamma}$ is the loss time due to pion production in inelastic $p\gamma$ collisions on the high-energy photons emitted by the CS. For $\tau_{\rm p\gamma}$, we calculate here a conservative (and pessimistic) lower value, $\tau_{\rm p\gamma, min}$, which assumes that the CS radiates all the energy it processes in photons with energies set to the threshold for pion production.

\section{Results}
\label{Sec_Results}

\begin{figure*}
\begin{center}
\includegraphics[width=0.49\textwidth]{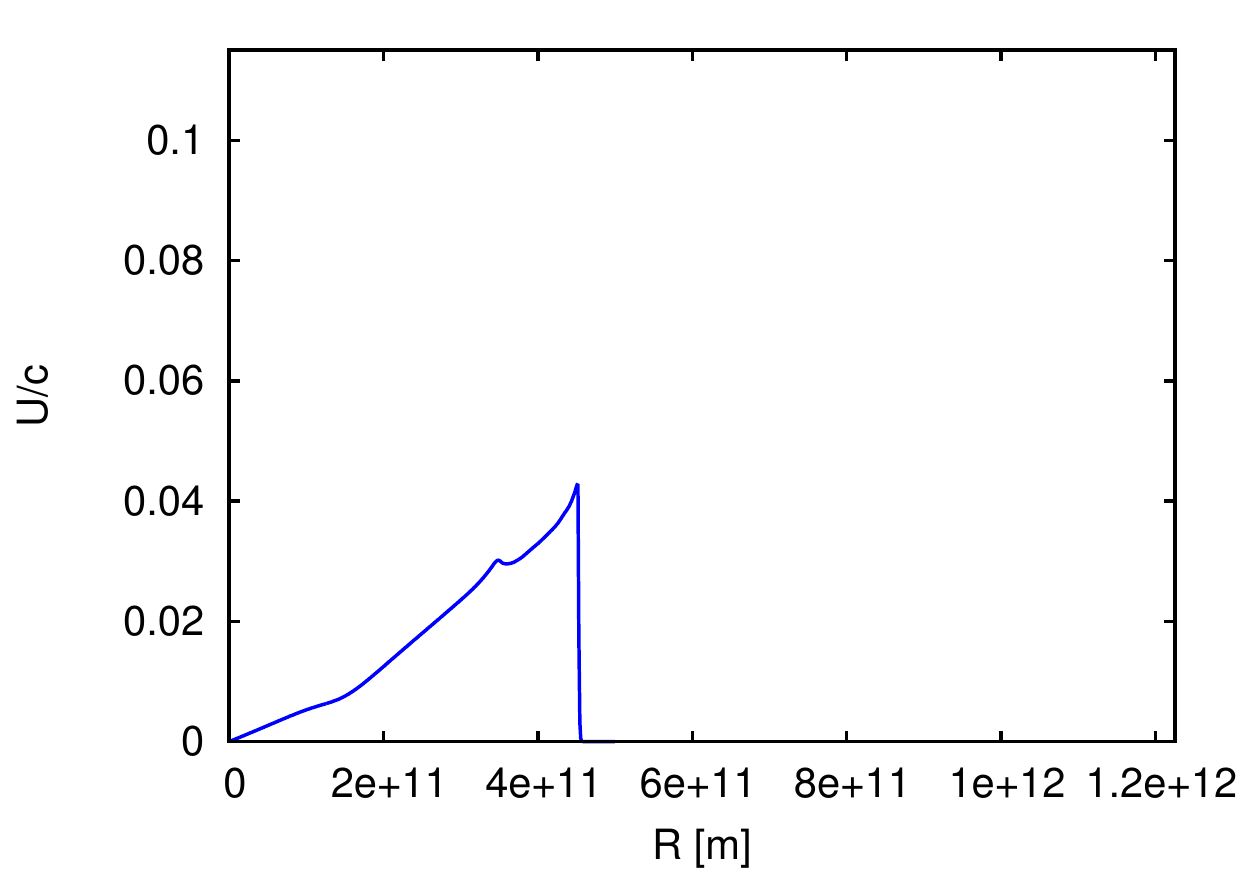}
\includegraphics[width=0.49\textwidth]{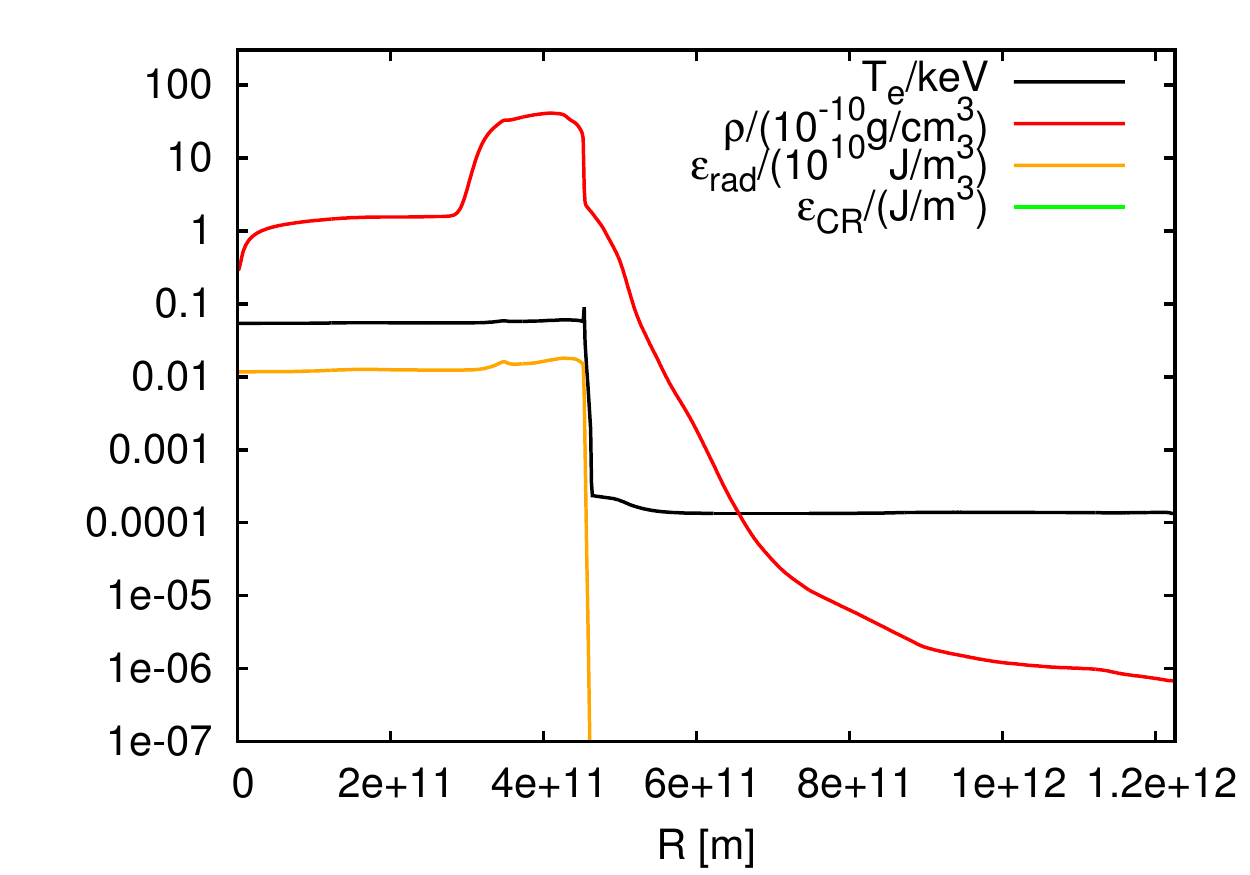}
\includegraphics[width=0.49\textwidth]{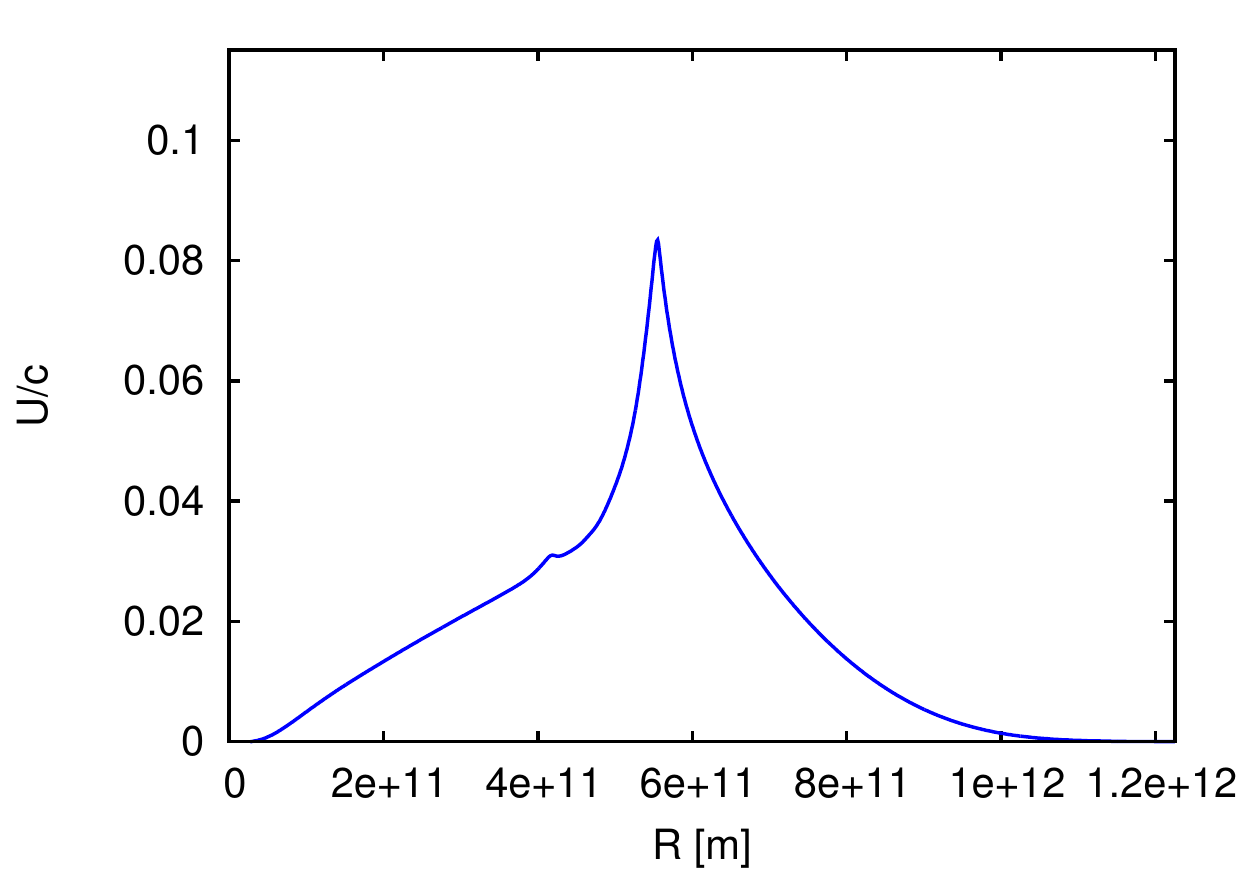}
\includegraphics[width=0.49\textwidth]{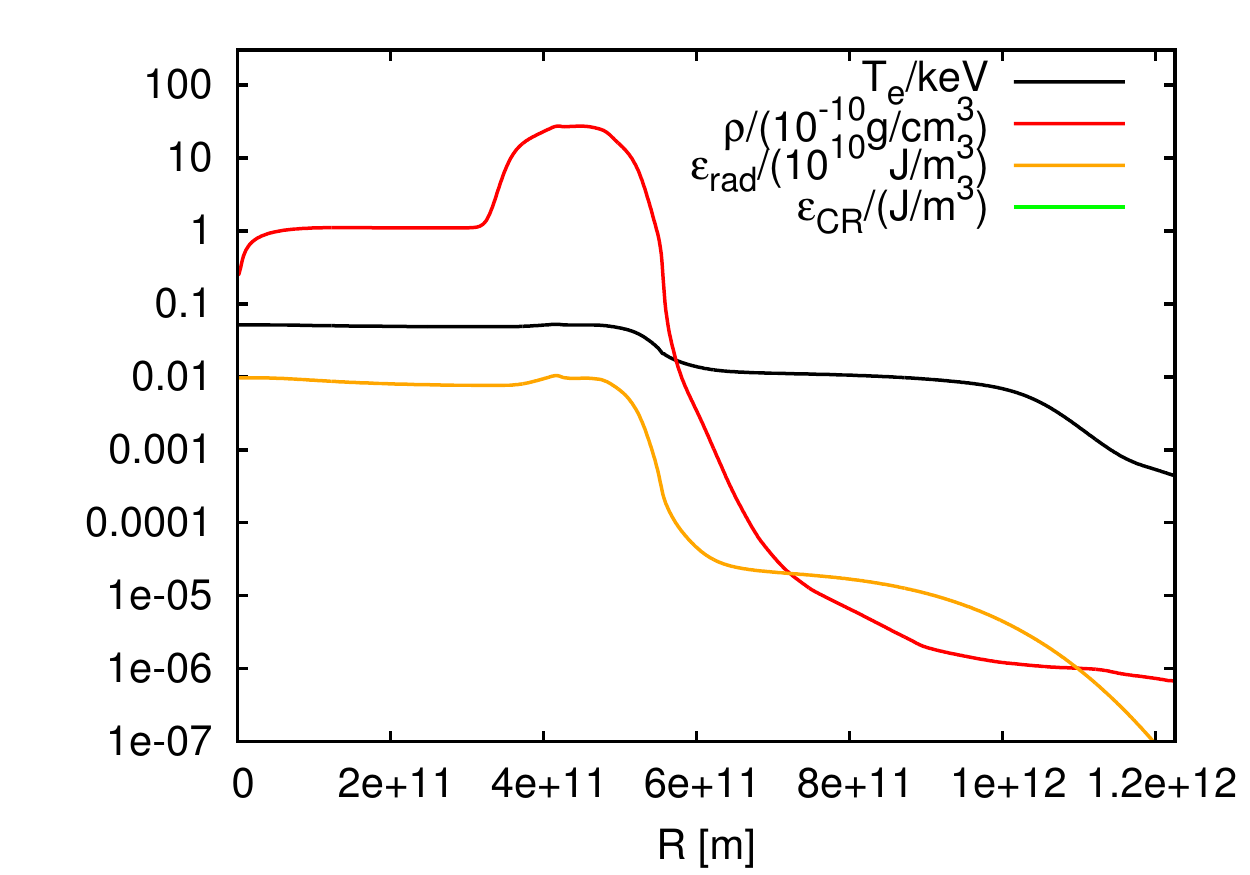}
\includegraphics[width=0.49\textwidth]{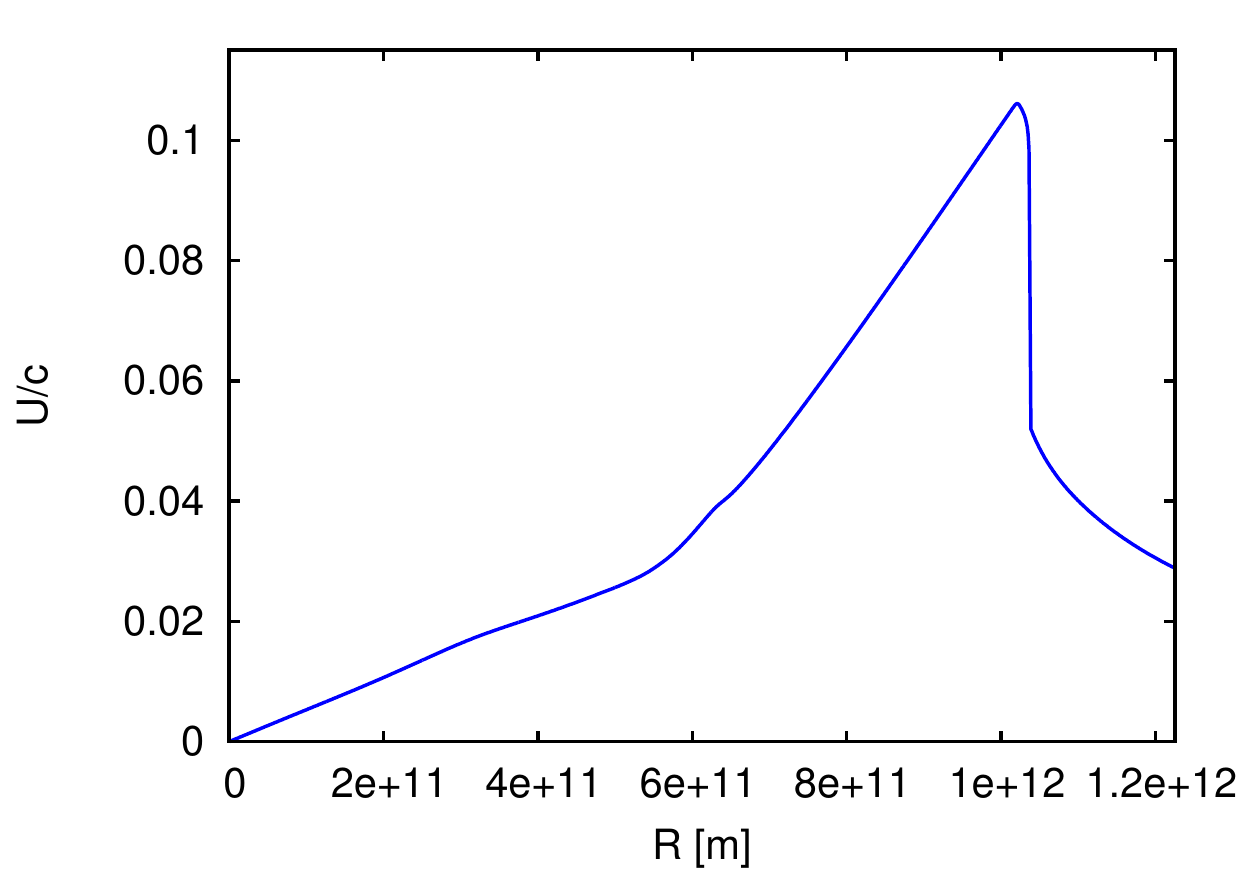}
\includegraphics[width=0.49\textwidth]{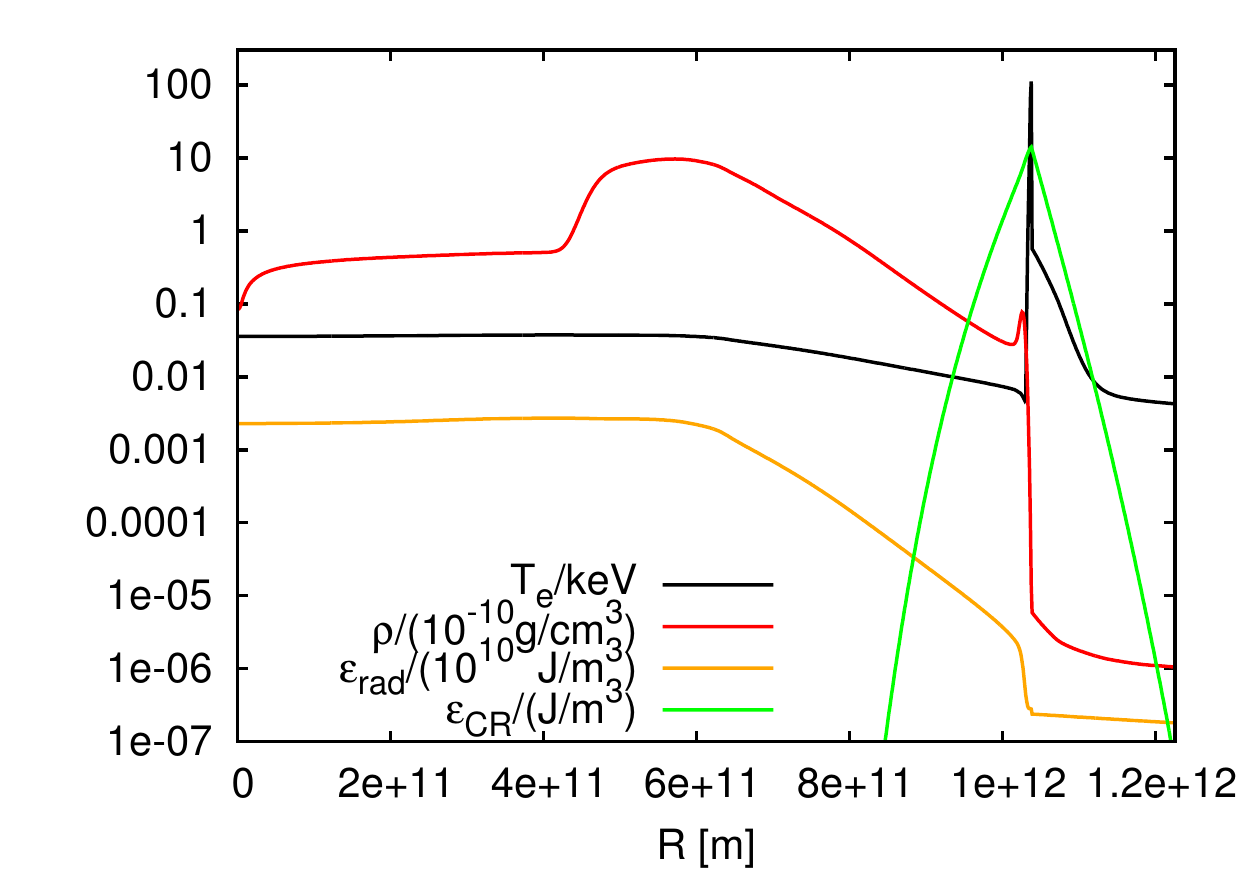}
\caption{\label{Fig2} Simulation of the explosion of a red supergiant. Velocity of the fluid normalized to $c$ (left column), and electron temperature $T_{\rm e}$, fluid density $\rho$, radiation and CR energy densities $\eps_{\rm rad,CR}$ (right column) at three different times: shortly before (upper row), during (middle row) and soon after (lower row) SB ---the optical depth is equal to 1 at $R \sim 5.5 \cdot 10^{11}$\,m in the initial profile. See the keys in the right column for the line colours and normalizations.}
\end{center}
\end{figure*}

We present in Fig.~\ref{Fig2} results from our CR-radiation-hydrodynamics simulation of the explosion of the star, at three different times: Shortly before SB in the upper row, during SB in the middle row, and soon after SB in the lower row. In the left column, we show $U/c$, the velocity of the fluid normalized to $c$, as a function of the radius $R$, counted from the centre of the star. In the right column, we plot the electron temperature $T_{\rm e}$ (black lines), the fluid density $\rho$ (red), the radiation energy density $\eps_{\rm rad}$ (orange), and the CR energy density $\eps_{\rm CR}$ (green), as functions of $R$. See the keys in the Figure for the units and normalizations. In the initial profile, the optical depth is equal to 1 at $R \sim 5.5 \cdot 10^{11}$\,m. In the first row, the shock is still inside the optically thick layers of the star, and is radiation-mediated. It is located at $R \simeq 4.5 \cdot 10^{11}$\,m, see the vertical jump in the blue curve in the left panel. The shock thickness is several photon mean free paths, and the post-shock radiation is well confined, see the orange line in the right panel. In the middle row, the shock reaches the outer layers of the star. The radiation in the region immediately behind the shock starts to escape, and creates a flash of photons: The inflection around $R \simeq 1 \cdot 10^{12}$\,m in the curve for the radiation energy density (orange line in the right panel) corresponds to the front of SB photons which escape at the speed of light towards $R \rightarrow \infty$. The escaping radiation accelerates the circumstellar medium, which is visible in the left panel: The discontinuity in $U/c$ is now significantly wider, and spans from $R \simeq 6 \cdot 10^{11}$\,m to $R \simeq 10^{12}$\,m. In the lower row, the CS has formed: In the left panel, it appears as the abrupt discontinuity in $U/c$ at $R \simeq 1.05 \cdot 10^{12}$\,m, embedded within a significantly broader transition region smoothed by the radiation from SB. One can see in the right panel a spike in temperature ($\sim 100$\,keV), which corresponds to the shock-heated region in the downstream of the newly formed CS. CR acceleration has started at the CS, cf. the green curve.

\begin{figure*}
\begin{center}
\includegraphics[width=0.49\textwidth]{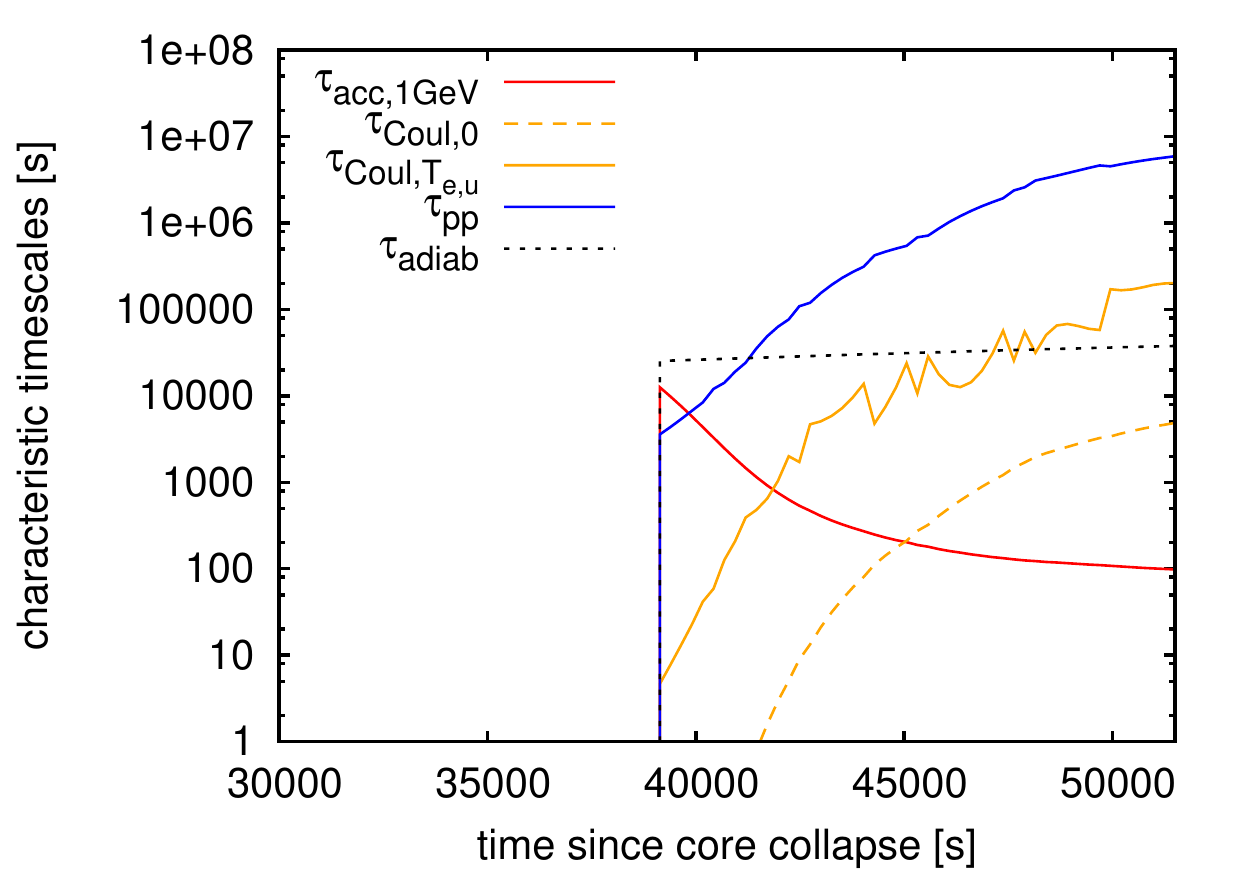}
\includegraphics[width=0.49\textwidth]{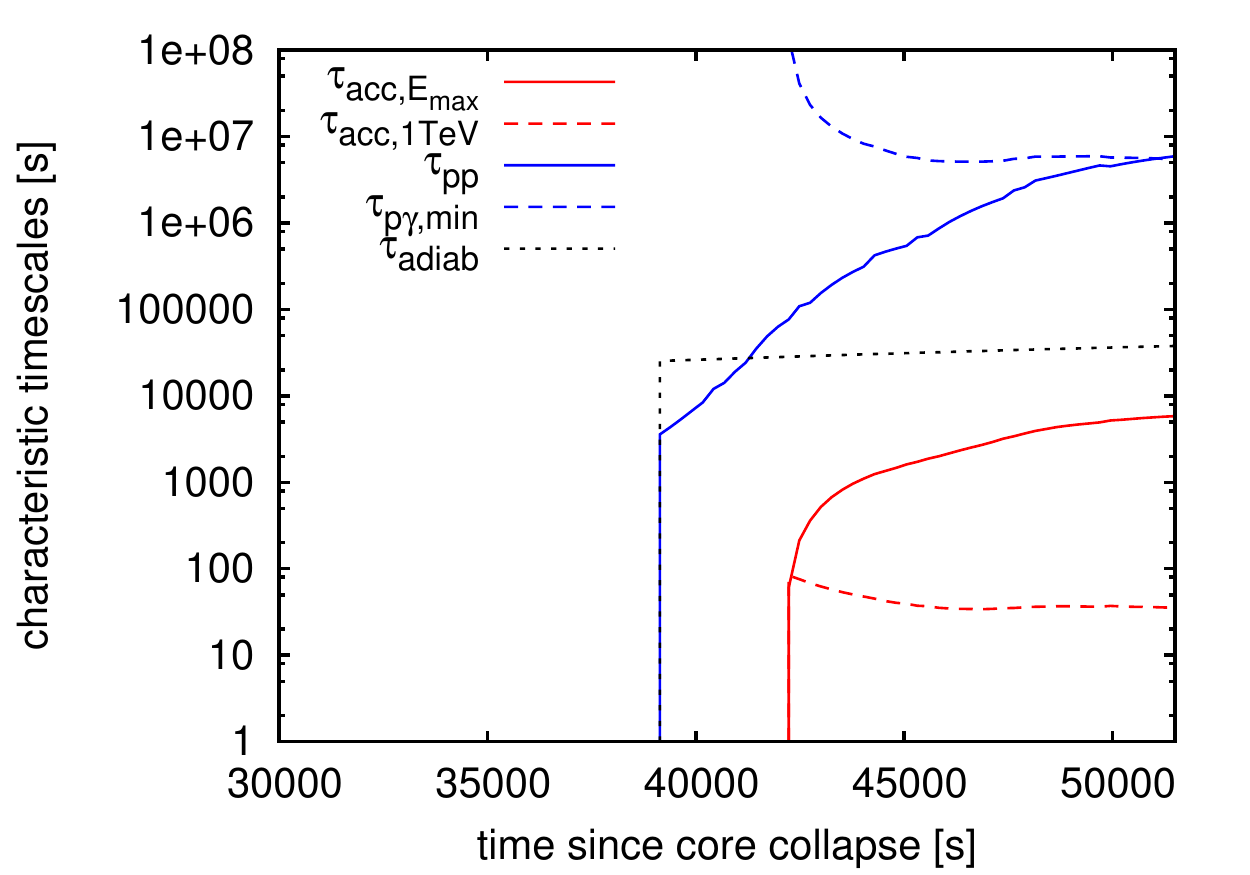}
\caption{\label{Fig3} Characteristic timescales for particle acceleration and energy losses, versus time since core collapse. See the keys for the line types and colours, and see the text for more details.}
\end{center}
\end{figure*}

We show in Fig.~\ref{Fig3} our calculations of the typical timescales for particle acceleration and energy losses, as functions of time since core collapse. In the left panel, the solid red line corresponds to $\tau_{\rm acc, 1GeV}$, the CR acceleration time to 1\,GeV in an initial circumstellar magnetic field set to 1\,mG. It is plotted against the loss time for inelastic $pp$ collisions ($\tau_{\rm pp}$, solid blue line), as well as the adiabatic ($\tau_{\rm adiab}$, dotted black), and Coulomb loss times using the electron temperature in the CS upstream from the simulation ($\tau_{\rm Coul, T_{\rm e,u}}$, solid orange). For information, we also show the limiting Coulomb loss time one would have assuming that the electron temperature is zero ($\tau_{\rm Coul, 0}$, dashed orange). This plot shows that Coulomb losses are the limiting factor for the onset of CR acceleration here. In this simulation, the CS forms around $t \simeq 39000$\,s, and the red line drops below the solid orange one around $t \simeq 42000$\,s. This shows that Coulomb losses may delay the onset of CR acceleration after the formation of the CS by only $\sim 1$\,hour for this progenitor and a 1\,mG circumstellar magnetic field. In the right panel of Fig.~\ref{Fig3}, we plot the CR acceleration time to $E_{\max}$ ($\tau_{\rm acc, E_{\max}}$, solid red line) and to 1\,TeV ($\tau_{\rm acc, 1TeV}$, dashed red) in the amplified magnetic field and assuming Bohm diffusion. We plot it against the $pp$ and adiabatic loss times, as well as the minimum $p\gamma$ loss time for CRs with energy $E_{\max}$ ($\tau_{\rm p\gamma, min}$, dashed blue line). Fig.~\ref{Fig3} (right panel) demonstrates that CR acceleration to $E_{\max}$ is not hindered by any of these losses.

\begin{figure*}
\begin{center}
\includegraphics[width=0.49\textwidth]{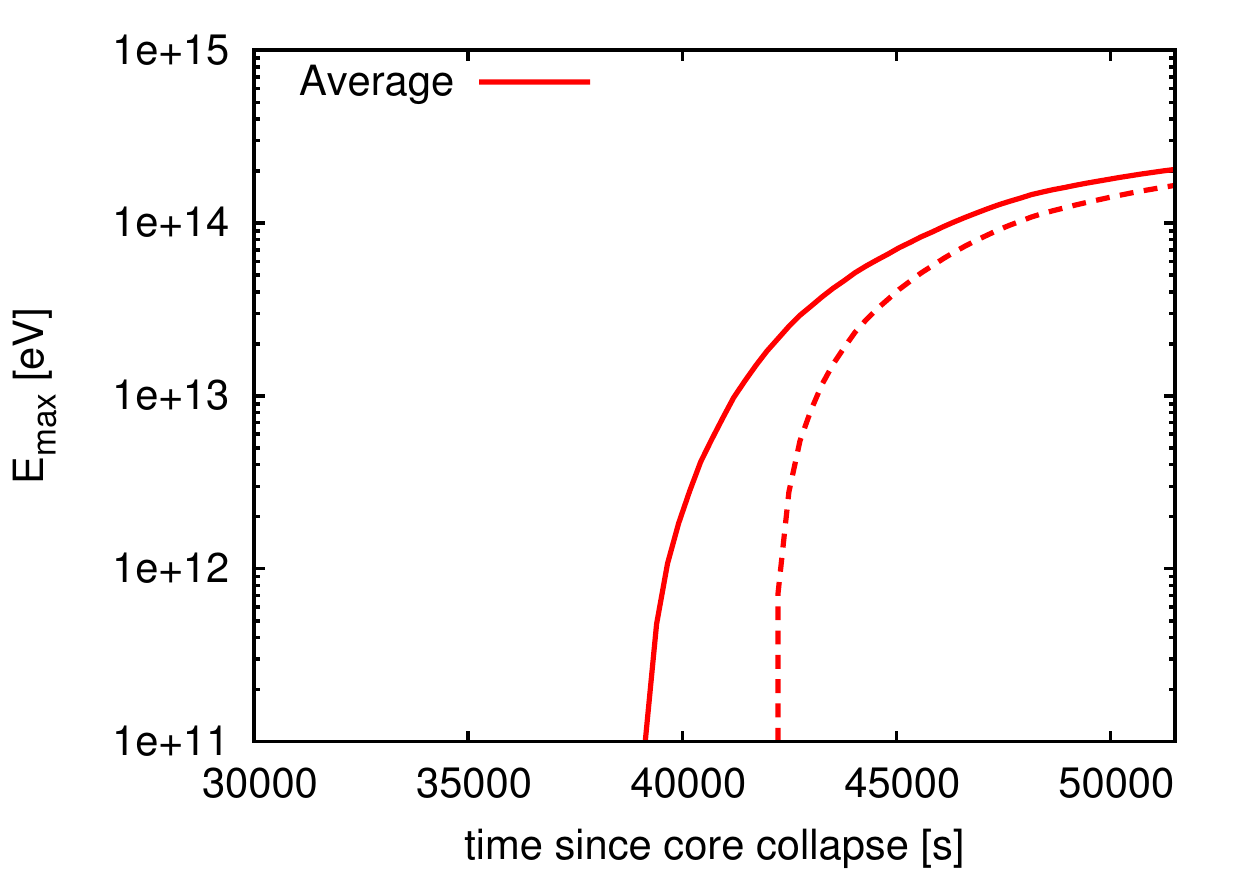}
\includegraphics[width=0.49\textwidth]{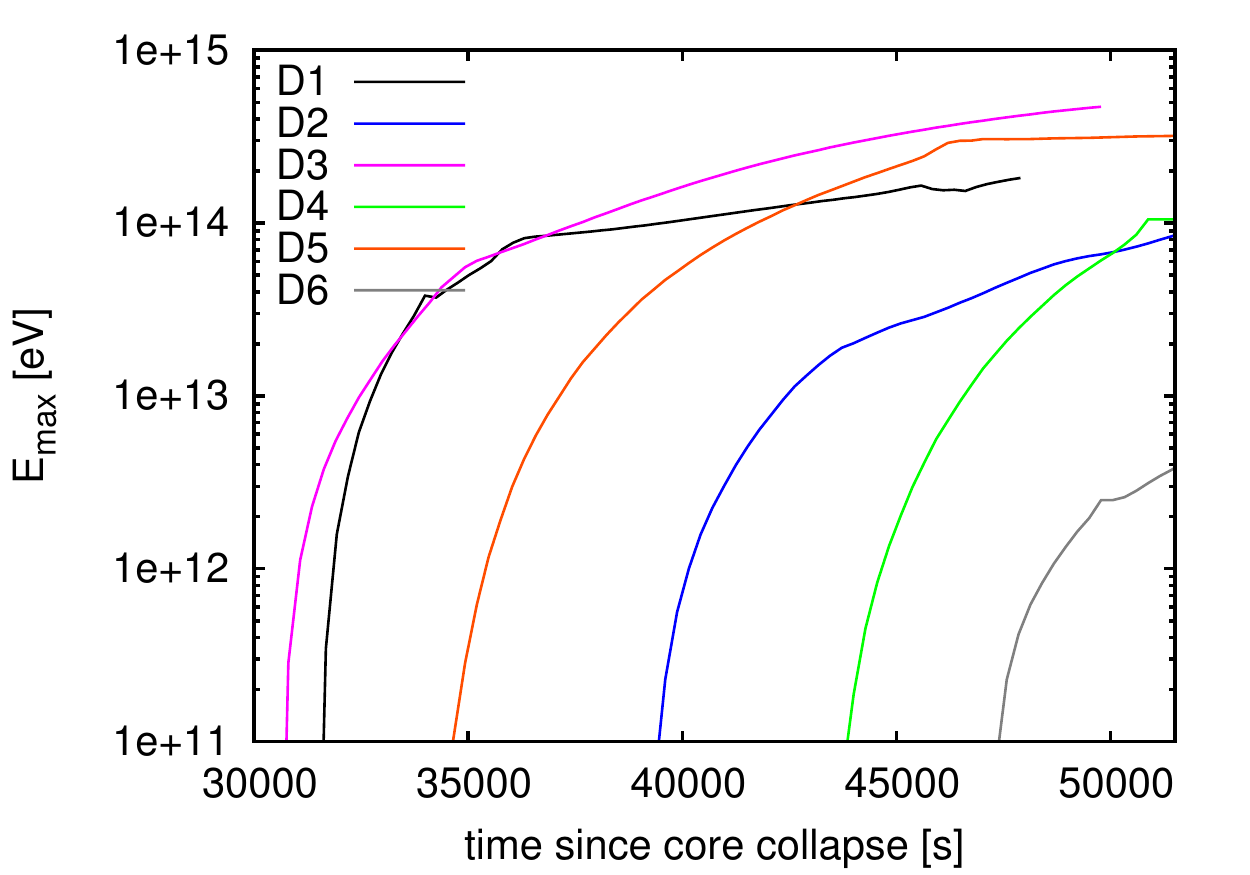}
\caption{\label{Fig4} Maximum CR energy, $E_{\max}$, versus time since core collapse, for the initial (circum)stellar density profile averaged over all radial directions (left panel) and for the initial density profiles along six given radial directions, \lq\lq D1\rq\rq\/ to \lq\lq D6\rq\rq\/ (right panel). In the left panel, the solid line corresponds to negligible Coulomb losses, and the dashed line assumes that CR acceleration starts when $\tau_{\rm acc,1GeV} \leq \tau_{\rm Coul,T_{\rm e,u}}$.}
\end{center}
\end{figure*}

We plot in Fig.~\ref{Fig4} our calculations of the maximum CR energy at the forward shock, $E_{\max}$, versus time since core collapse. Results in the left panel are for a simulation in the average density profile of the SN progenitor and its surroundings. The solid red line is for the optimistic case of CR acceleration starting as soon as the CS is formed, i.e. for negligible Coulomb losses. This may be possible if the magnetic field strength around the progenitor star is $\gg 1$\,mG. The dashed line assumes that particle acceleration starts when $\tau_{\rm acc,1GeV} \leq \tau_{\rm Coul,T_{\rm e,u}}$. These results show that CRs reach $\geq 100$\,TeV energies typically $\sim 1$\,hour afer the onset of particle acceleration, and $E_{\max}$ starts to plateau to values of a few hundreds of TeV by the end of the first day following core collapse. The fact that both lines reach almost the same limiting value shows that Coulomb losses have little impact on $E_{\max}$, and only slightly delay the beginning of particle acceleration. The results in the right panel are for six density profiles along six given radial directions, denoted as \lq\lq D1\rq\rq\/, \lq\lq D2\rq\rq\/, ...,\lq\lq D6\rq\rq\/. The profiles D1 to D4 correspond to those plotted in Fig.~\ref{Fig1} (right panel) ---the profiles D5 and D6 are not shown there due to limited space in the plot. The diversity of the curves in Fig.~\ref{Fig4} (right panel) provides an estimate of how the 3D geometry of the progenitor affects the results for $E_{\max}$ versus time. First, the times at which the CS forms and CR acceleration starts depend on the considered direction around the star. Indeed, the progenitor is not perfectly spherically symmetric, which implies that SB and thence the onset of CR acceleration occurs earlier in some directions than in others. However, we caution that a 3D simulation would be needed for a more watertight estimate of the spread in SB times: In 3D, the forward shock can get around dense clumps, which cannot be taken into account in our simulation. Second, the limiting value of $E_{\max}$ at which each curve plateaus depends on the direction. This means that the limiting $E_{\max}$ is not exactly the same in each region of the forward shock, at least during the first day of the SN. For example, the magenta curve reaches a value that is a few times larger than that of the black curve. Looking at Fig.~\ref{Fig1} (right panel), one can see that the circumstellar material at $R \geq 6 \cdot 10^{11}$\,m is about an order of magnitude denser in the direction D3 (magenta curve) than in the direction D1 (black one). These results are in line with the calculations of Ref.~\cite{Bell:2013kq}, where $E_{\max}$ is found to be larger in denser regions ---at $U_{\rm s}$ fixed. All six curves nonetheless tend towards the same order-of-magnitude values of $E_{\max}$, which shows that taking the average density profile provides a reasonable first estimate.

\section{Conclusions}
\label{Sec_Conclusions}

We study here the formation of a CS and the beginning of CR acceleration at a Type II SN, during the first day that follows core collapse. We present the CR-radiation-hydrodynamics simulation of the explosion of a red supergiant with a realistic density profile. We find that a CS soon forms after SB, and that CR acceleration starts soon after. For the progenitor considered here, Coulomb losses do not delay CR acceleration by more than an hour. The maximum CR energy at the forward shock quickly increases, and already reaches a few hundreds of TeV only hours after SB. It starts to plateau towards its maximum value by the end of the first day of the SN. The density of the circumstellar medium in the immediate surroundings of the progenitor depends on the considered radial direction around the star, and we find that CRs can be accelerated to higher energies in denser regions. Our results suggest that core-collapse SNe are plausible sources of very high-energy CRs, even in their earliest phases.

\acknowledgments

We thank Matthieu Renaud, Vincent Tatischeff, and Pierre Cristofari for useful discussions. This research collaboration is supported by a grant from the FACCTS program to the University of Chicago (PI: VVD). This work is supported by the ANR-14-CE33-0019 MACH project.



\end{document}